\documentclass[
]{ceurart}
\begin{document}

\copyrightyear{2021}
\copyrightclause{Copyright for this paper by its authors.
  Use permitted under Creative Commons License Attribution 4.0
  International (CC BY 4.0).}

\conference{QSET'21 - 2nd Quantum Software Engineering and Technology Workshop
co-located with IEEE International Conference on Quantum Computing and Engineering (QCE21) (IEEE Quantum Week 2021)
October 18–22, Virtual Conference}

\title{Cybersecurity for Quantum Computing}

\author[1,2,3]{Natalie Kilber}[%
orcid=0000-0001-6354-496X,
email=natalie.kilber@nablaco.com,
url=https://www.nablaco.com/,
]
\address[1]{MHP Management- und IT-Beratung GmbH, Königsallee 49, Ludwigsburg 71638, Germany}
\address[2]{Nablaco - A Natalie Kilber Advisory, Siemensstrasse 3, 71696 Moglingen, Germany }
\address[3]{Institute of Software Engineering - University of Stuttgart,
  Universitätsstrasse 38, 70569 Stuttgart, Germany}
\author[4,5]{Daniel Kaestle}[
]
\address[4]{Member of the Cyber Security Sharing and Analytics Association (CSSA)}
\address[5]{Certified Information Systems Security Professional (CISSP)}

\author[3]{Stefan Wagner}[%
orcid=0000-0002-6799-6709,
email=stefan.wagner@uni-stuttgart.de,
url=https://www.iste.uni-stuttgart.de/institute/team/Wagner-00017/,
]
\begin{abstract}
With rising cyberattack frequency and range, Quantum Computing companies, institutions and research groups may become targets of nation state actors, cybercriminals and hacktivists for sabotage, espionage and fiscal motivations as the Quantum computing race intensifies. Quantum applications have expanded into commercial, classical information systems and services approaching the necessity to protect their networks, software, hardware and data from digital attacks. This paper discusses the status quo of quantum computing technologies and the quantum threat associated with it. We proceed to outline threat vectors for quantum computing systems and the respective defensive measures, mitigations and best practices to defend against the rapidly evolving threat landscape. We subsequently propose recommendations on how to proactively reduce the cyberattack surface through threat intelligence and by ensuring security by design of quantum software and hardware components. 
\end{abstract}

\maketitle
\section{Introduction}
Cybersecurity is preparing for what is next, yet it is often an afterthought. With the cybersecurity breaches on the rise\cite{gtr}, academia, companies, institutions and Quantum Computing groups may become targets for cybercriminals Hacktivist and Advanced Persistence Threats as the Quantum Computing race intensifies. Quantum applications have expanded into classical information systems approaching the motivations and needs of state of the art Software Engineering practices. Thus it is an imperative to build in security during the design phase and therefore shift security left to ensure Quantum Computing ecosystems, services, as well as promising technologies are reliable and secure to use. \\

Continuous technological advancements towards a digital ecosystem encompass laboratory environments, businesses, institutions, operational technologies and every connected device. The recent shift to remote work and the growing digitalization is paralleled by the proliferation of cyberattacks. Entities collecting and storing sensitive data such as intellectual property regardless of sector or size are at a higher risk of being targeted for a cyberattack, be it espionage or sabotage. Accruing attacks on supercomputers, academia, research and development sites as of late \cite{archer,julich,uk} underline the premise. 
Various aspects of Quantum Computing Security have been covered by many surveys and other papers \cite{ncc,pqc1,pqc2,qriskcatas}. The main idea of this work covers an overarching topic across every industry with digital capabilities to academic institutions and groups, developing, designing or working with Quantum Computing Software or Hardware including Quantum Technologies connected to a network. Startups, university spin-offs and vendors benefit from a discussion early on if they introduce secure proprietary technologies for users and secure their solutions from relevant threats. This survey’s intent also lies in spurring discussions and furthering ideas in interdisciplinary research to reflect on educational paths for Quantum Computing specialists.\\
Along with this focus, this paper will make a distinction between quantum computing systems and quantum technologies with different technological maturities. The latter range spans wide and presents higher technological sophistication such as quantum - communication, - metrology and - sensing, and the simulation and numerical techniques associated with it \cite{qtflag}.

\section{Preliminaries}
Before getting to the threat vectors, we will review a few topics pertinent to understanding the capabilities of quantum computers. Quantum Computing systems bring a number of components together to form an optimized specialized processor for utilizing quantum phenomena in their computation. Amdahl’s Law determines a co-processor linked to a central processing unit capable of speeding up the overall execution, respectively of specific computational intensive kernels, to be an accelerator \cite{MLsurvey}. To make a quantum computing system Turing complete, a quantum processing unit (QPU), the co-processor, requires an analog to digital interface to convert analog signals back and forth between the control system, which in turn requires for the application logic a Host-CPU that may connect to a network \cite{quaccel}. Figure 1 captures an example of a general quantum accelerator architecture. For instance, a Quantum as a Service, QaaS, offers managed compute capacity on demand over the cloud or any other remote access.\\

\begin{figure}
  \centering
  \includegraphics[width=\linewidth]{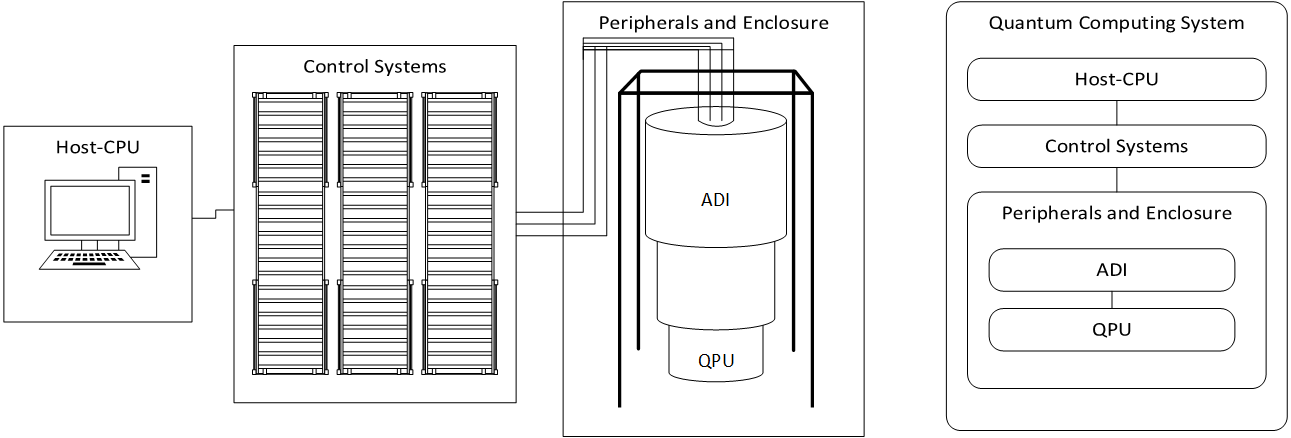}
  \caption{ A Quantum Computing Architecture}
\end{figure}

Underpinning these systems are the need for innovations in modern computing systems due to the stagnation of trends of technological improvements in clock frequency, instruction per Joule and clock cycle, core counts, transistor size and power density related to Moore’s, Koomey’s and Denard’s scaling laws\cite{Moore,Horowitz, MLsurvey}. Advancements are progressing, however, there are fundamental limits that these specialized circuits including QPUs have to abide, such as the Landauer limit \cite{Landauer}. Given the low maturity of these systems, downtimes for calibration and maintenance are imperative and elaborate. Existing bottlenecks in quantum computing are often tied to unresolved fundamental research questions such as missing solutions for physically realisable Quantum random access memory (QRAM) \cite{qram}, as commercial QPUs operate to date with read-only-memory (QROM). Another unsolved problem, rendering today's devices under the Noisy-intermediate-scale Quantum (NISQ) era \cite{Preskill} is cross-talk error \cite{crosstalk} given limited error correction schemes and low fidelity - the reliability of the computation given the noise of quantum gates. Open questions remain about the scaling to reduce the size of the devices, especially wiring for the control reaching into cryogenic fridges i.e. for superconducting quantum chips \cite{msft}, although some types of photonic quantum chips can operate in ambient temperatures, which again have a scaling problem due to bulky optical components. Other bottlenecks include the necessity for exotic materials such as rare earth  metals for the fabrication of components and cryogenic environments having to be run on Helium, which is a very limited resource to name a few unsolved supply chain constrictions.\\

Further on, we will discuss the types of applications for which these quantum accelerators are being designed in order to understand the value and motivation behind them. The flood of news and articles about how Quantum Computing can break current security instances in the future skews the picture of the status quo and misses to convey that these potential capabilities are years away. Molecular simulation, catalyst or drug design and several optimization problems are just a few applications a reliable quantum accelerator might conceivably push scientific discoveries further \cite{Preskill}. Albeit, such systems have yet to be designed and so classical counterparts dominate these optimizations and advancements for now. Take, for example, the case of theoretical conjectures about RSA-2048 bit decryption with quantum accelerators. Recent work by Google estimate 20 million NISQ device qubits to crack an RSA key within 8h \cite{google}, whilst the newest developments in physical - not error corrected logical - qubits have not officially surpassed the 100 qubit mark for NISQ devices \cite{IBM}, excluding digital annealer technologies - special purpose analog machines for optimization. Other theoretical advances promise the possibility of cracking RSA-2048 encryption with  13436 qubits in 177 Days \cite{rsa177} and the premise of a multimode quantum memory, which also is a theoretical presumption that has not been found. Even if we had a quadratic speedup due to a Grover algorithm for quantum accelerated pre-sampling to brute force key search against AES, it wouldn’t be possible to break anything above AES-256bit key length with supercomputers nowadays \cite{Grover}. Here, researchers argue that classified, encrypted data with longer intelligence life and shorter key length than AES-256, can be stored and broken in the future \cite{crypto}. These may face a quantum threat from the future, but intelligence with a shorter lifetime is not affected given that decryption might take months. Hence, the reverse threat of quantum computing devices are not imminent and the intelligence life of data plays a significant role in it.\\

Yet, the issue of cybersecurity in nascent Quantum Computing resources is rarely discussed. As Quantum Computing systems are and will be hybrid systems for the foreseeable future with CPU-hosts, cloud-based or managed APIs, the need for reliable, secure services and architectures arises. Subsequently, the critical applications and data these systems will handle and store, demand a focus on appropriate security controls. 

\section{Threat Landscape}

Behind every cyberattack stands a threat actor and a motivation associated with it. Threat actors distinguish themselves through their objectives: politically motivated adversaries with nation state vicinity  are distinguished as nation-state actors, while financially motivated threat actors classify under cybercrime, hacktivists are driven by ideology, but the boundaries are blurry as some nation-state threat actors also have financial motives. The growing global importance of digitalisation and connectivity has also advanced the growth of the cybercrime ecosystem with many criminal enterprises supporting threat actors with complimentary data exfiltration, ransomware and malware-as-a-service operations. The ecosystem evolved to holistically cover cybercrime services: recruiting people, developing web-injection kits or exploit networks, specialized distribution mechanisms like spam email delivery, offering monetization schemes such as wire fraud and cryptocurrency services \cite{gtr}.\\

The race to scientific advancements in quantum computing has been coined to be of national security interest \cite{pqc1,pqc2}. Researchers suggest in overstated cryptographic claims that "the nation that wins the quantum race will be able to protect their secrets with a higher level of security than contemporary cryptography guarantees and have unfettered access to those of the states that lost it" \cite{qriskcatas}. Irrespective of the exaggerations, the scientific race towards funding quantum computing and the subjacent hype is substantial for future applications of quantum computers claiming to solve high-margin economic problems previously intractable by conventional computational methods. Therefore the quantum race signifies an extension of state power and a promised economic advantage of first-movers \cite{crypto}. Accordingly, sabotage and research espionage lie at the core of nation-state adversary objectives. As availability and economic continuity of quantum services will mature the reliability, quality and responsibility of quantum computing providers, this will inspire attacks to take these systems down. In the same way, it will prompt cybercrime actors to seek financial gain in the sabotage, cyber extortion and data theft of these systems. Hacktivists may disapprove of the use of limited resources, the type of research being done or the mission statements of the targets seeking out similar tactics for their ideological motivations.\\
On an abstract level, adversary tactics are similar. The first step is reconnaissance to get to know the target, followed by establishing a foothold, escalating privileges and then proliferate throughout the network undetected. If the motive is espionage, the adversary will try to remain undetected and cause little disturbance, whereas if an attacker’s  motive is sabotage, disruption and damage follow unveiling the presence of an adversary. Financial gain objectives follow the same example coupled with either communication for extortion or data being sold off in the background.
To defend against specific threat vectors and threat actor strategies, their patterns of behaviour, bundled under the concept of Tactics, Techniques, and Procedures (TTPs) need to be known. Even more so, the organization or institution being targeted needs to know their own system and environment to effectively prevent possible attacks.\\

Quantum computing systems either run in enterprise ecosystems constituting of one or a mix of Windows, macOS, Linux, Azure AD, SaaS, IaaS, Network, Containers, etc. platforms; or they are part of industrial control systems (ICS) often managed by a Supervisory Control and Data Acquisition (SCADA) system, via programmable logic controllers (PLCs) or discrete process control systems (DPC) \cite{SANStarget}. Even in the case of ICS, the control systems are rarely, air-gapped, that is, physically separated from any network.\\

Targets for threat actors can be QaaS (Quantum-as-a-Serive), quantum application providers, as well as users processing or consuming such services. Top vulnerabilities for cloud, web applications and ICS systems apply. Common cloud security risks encompass the misconfiguration of services, infrastructure security, service or data integration and non-production environment exposure \cite{cloud}. While the prime web application security risks haven’t changed drastically over the last decade and OAWSP Top Ten \cite{OWASP} represent an ample base for minimizing these risks. Some of them are broken authentication methods or access control, sensitive data exposure or cross-site scripting, where an attacker might take advantage of an API or manipulate the DOM (Domain Object Model) to hijack user accounts, access browser histories, control browsers remotely or spread malware.
The most widespread types of vulnerabilities in ICS components differ slightly due to the hardware affinity. Older control systems are a source of many severe vulnerabilities and exploits against unpatched systems are widely available. OPC technology (Open Platform Communications) inherit the vulnerabilities associated with respective RPC (Remote Procedure Call) and DCOM services (Distributed Component Object Model), where a full range of ports are open to communicate by default, especially for older control systems like OPC classic. Buffer overflows, the use of hardcoded credentials and cross-site scripting are amongst the top three automated ICS vulnerabilities \cite{kaspersky, dragos}.
At last, arguably the most important aspect is the human component. 
Social engineering and phishing attacks have high rates of success exploiting human nature, i.e. associativity and curiosity. USB drop attacks, impersonation and asking for internals can lead to spear phishing and credential theft.
Not just legacy systems, new software design and development too, has its drawbacks and dangers.
The use of hard-coded credentials, missing or improper authorization or authentication for critical functions, incorrect default permissions and the exposure of sensitive information to unauthorized actors are part of the most common and most dangerous software weaknesses.  The Top 25 Most Dangerous Software Weaknesses scored over the last two years can be found on the CWE Top 25 List (Common Weakness Enumeration Top 25) \cite{cwe}. A more detailed view on TTPs of attackers for Enterprise Systems can be found on the MITRE ATT\&CK framework \cite{Mitreenter} or for the ICS systems on MITRE ATT\&CKICS, which is a common framework used by the cybersecurity industry \cite{Mitreics}.

\section{Defensive Measures, mitigations and best practices}

The first step is knowing your own technology stack and environment. 
Systems with a high level of cybersecurity maturity exhibit security operation programs with extensive logging and monitoring for threat detection and response activities. Dependent on the application, system or processes, there is always a trade-off between cost and available resources. Identifying relevant risks to your critical assets and processes aids in putting vital security controls in place.\\

Cyber hygiene and people awareness are your first line of defense you can easily strengthen. Remote work policies need to be in place as your attack surface expands into the cloud and homes’ of your employees. This means endpoints need to be protected, data at rest and in transit should be encrypted. Depending on the risk appetite and critical assets and applications, hybrid and cloud architectures need to be well configured regarding segmentation, authorization, authentication and encryption for relevant perimeters with firewalls and DMZs (Demilitarized Zone in perimeter networks) or follow a complete zero trust model if resources allow it. Cloud security gap analyses and security reviews help in finding misconfigurations and weaknesses. Non-production environments should not be neglected - especially research and development systems. As for legacy software, for instance in ICS systems, OPC classic should be migrated to OPC UA and further segmented to close the threat aperture with AAA (Authentication, Authorization and Accounting) and instantiate firewalls, DMZs, where it is sensible.
Quantum Computing Systems embrace a mix of off-the shelf components with proprietary software and hardware, which implicates the responsibility for flashing by design and releasing source-code bug fixes if for commercial use. The responsibility in off-the shelf components lies in recognizing vulnerabilities and patching them.\\

Having a closer look on control systems, fail-safe systems should be segmented thereby preventing single-points of failure. For instance, an attacker shouldn’t be able to have access to the cooling system and control unit of the ADI/QPU through the same Host-CPU for a sabotage attack on a compromised Quantum Computing System.
The people vector is also pivotal in ICS environments. An air-gapped system can only be breached by bridging that gap to gain physical access. Unauthorized access by an insider threat or unaware employee can be mitigated by tightly locking up physical access, only whitelisting approved USB sticks and/or having a device antivirus scan stage implemented. \\

A recommended set of actions for cyber defense in critical control systems can be found on SANS CIS Controls \cite{CIS} and for a comprehesive view on defense measures against sepcific tactics, visit MITRE Shield \cite{MitreD}. OPC ICS Security Tools help design for more secure OPC systems \cite{ICStool}.
OWASP’s CLASP \cite{clasp} and Microsoft’s SDL \cite{SDL} help in identifying security vulnerabilities during every software design and development phase to have security built into the product. The DevSecOps framework goes a step further in automating the integration of security tools and processes in every phase of the software development lifecycle. Tools such as Metasploit or W3AF allow developers to robustly test for any potential vulnerabilities \cite{pratte}.

\section{Conclusion}

Given the limited amount of resources and funding for quantum computing research and fewer devices commercially available, we shouldn’t wait until something goes wrong. For that reason alone, we should build and design quantum hardware and software components with security in mind. Equally, the design of corresponding services should be shifted left.
The Quantum Computing industry is a likely target for sabotage, espionage and extortion motives. Threat intelligence could change the otherwise reactive security activities to a more proactive fight against threat actors and secure one’s system with foresight. MISP (Malware Information Sharing Platform) is an open source threat intelligence platform \cite{misp} and a good place to start understanding the types of threat actors.

\section{Acknowledgements}

We thank Pradeep Vingesh Raghupathy and  Helmut G. Katzgraber for fruitful discussions. The views and conclusions contained herein are those of the authors and should not be interpreted as necessarily representing the official policies or endorsements, either expressed or implied, of the CSSA.

\bibliography{csqc}

\end{document}